\documentclass[11pt]{article} 
\usepackage[a4paper,margin=2.5cm]{geometry}

\usepackage[T1]{fontenc} %
\usepackage[normalem]{ulem} %

\usepackage[usenames,dvipsnames,svgnames,x11names]{xcolor}

\usepackage[nocompress]{cite}

\usepackage{balance}

\usepackage{listings}

\lstdefinelanguage{XML}
{
basicstyle=\ttfamily\footnotesize,
  morestring=[b]",
  moredelim=[s][\bfseries\color{Maroon}]{<}{\ },
  moredelim=[s][\bfseries\color{Maroon}]{</}{>},
  moredelim=[l][\bfseries\color{Maroon}]{/>},
  moredelim=[l][\bfseries\color{Maroon}]{>},
  morecomment=[s]{<?}{?>},
  morecomment=[s]{<!--}{-->},
  commentstyle=\color{gray},
  stringstyle=\color{blue},
  identifierstyle=\color{red}
}

\usepackage{moreverb}

\usepackage[nounderscore]{syntax}

\usepackage[pdftex]{graphicx}
\graphicspath{{./figures/}}
\DeclareGraphicsExtensions{.pdf}

\usepackage[cmex10]{amsmath}
\usepackage{amssymb}
\usepackage{mathtools}
\usepackage{amsthm}
\usepackage{amsfonts}
\usepackage{gensymb}

\usepackage{subfig} %

\usepackage{algorithmicx}
\usepackage{algpseudocode}
\usepackage[ruled]{algorithm}
\definecolor{light-gray}{gray}{0.75}
\algrenewcommand{\algorithmiccomment}[1]{\hskip3em{{\footnotesize \textcolor{light-gray}{$\blacktriangleright$}}} #1}

\usepackage{multirow} %
\usepackage{rotating} %
\usepackage{booktabs} %
\usepackage{colortbl} %
\usepackage{tablefootnote} %

\usepackage{array}
\newcolumntype{L}[1]{>{\raggedright\let\newline\\\arraybackslash\hspace{0pt}}m{#1}}
\newcolumntype{C}[1]{>{\centering\let\newline\\\arraybackslash\hspace{0pt}}m{#1}}
\newcolumntype{R}[1]{>{\raggedleft\let\newline\\\arraybackslash\hspace{0pt}}m{#1}}

\usepackage[pdftex,colorlinks=true,urlcolor=blue,citecolor=blue]{hyperref}

\usepackage{xspace}

\usepackage{enumitem}

\usepackage{blindtext}

\newlength{\myeqskip}  
\AtBeginDocument{%
    \setlength\abovedisplayskip{\myeqskip}%
    \setlength\belowdisplayskip{\myeqskip}%
    \setlength\abovedisplayshortskip{\myeqskip-\baselineskip}%
    \setlength\belowdisplayshortskip{\myeqskip}}

\def\orcid#1{\kern .08em\href{https://orcid.org/#1}{\includegraphics[keepaspectratio,width=0.7em]{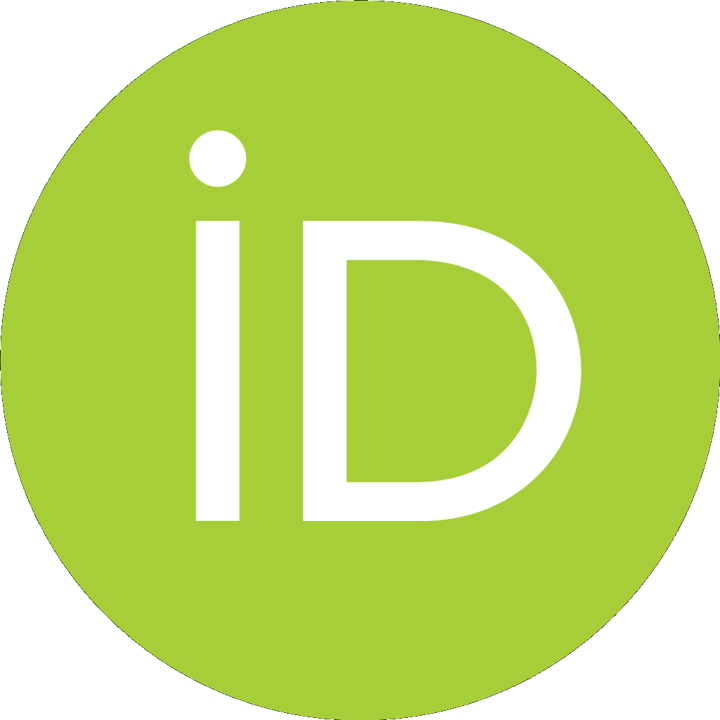}}}

\begin{document}

\title{
Towards AI Agents for Course Instruction in Higher Education:\\
Early Experiences from the Field}
\author{Yogesh Simmhan\orcid{0000-0003-4140-7774} and Varad Kulkarni\\
\small Department of Computational and Data Sciences,\\
\small Indian Institute of Science (IISc), Bangalore 560012 India\\
\small Email: \href{mailto:simmhan@iisc.ac.in}{simmhan@iisc.ac.in}}

\maketitle

\pagestyle{plain}

\begin{abstract}
This article presents early findings from designing, deploying and evaluating an AI-based educational agent deployed as the primary instructor in a graduate-level Cloud Computing course at IISc. We detail the design of a Large Language Model (LLM)-driven Instructor Agent, and introduce a pedagogical framework that integrates the Instructor Agent into the course workflow for actively interacting with the students for content delivery, supplemented by the human instructor to offer the course structure and undertake question--answer sessions.
We also propose an analytical framework that evaluates the Agent--Student interaction transcripts using interpretable engagement metrics of topic coverage, topic depth and turn-level elaboration. 
We report early experiences on how students interact with the Agent to explore concepts, clarify doubts and sustain inquiry-driven dialogue during live classroom sessions. 
We also report preliminary analysis on our evaluation metrics applied across two successive instructional modules that reveals patterns of engagement evolution, transitioning from broad conceptual exploration to deeper, focused inquiry. These demonstrate how structured integration of conversational AI agents can foster reflective learning, offer a reproducible methodology for studying engagement in authentic classroom settings, and support scalable, high-quality higher education.
\end{abstract}

\section{Introduction}
Conversational AI tools are increasingly being integrated into classrooms, offering new ways for students to learn through inquiry-driven and interactive engagement~\cite{abukhurma2024chatgpt}. Recent studies show that 
Large Language Models (LLMs) powered agents can enhance classroom participation, promote active inquiry, and support personalized learning experiences by providing contextual explanations and instant feedback~\cite{abukhurma2024chatgpt, chiu2023ai}. They allow students to engage in self-directed and personalized exploration while receiving adaptive, on-demand guidance from AI tutors.

Despite the growing interest and gradual adoption of AI agents in the classroom, few studies have examined how to design and incorporate such AI tutoring systems actively in a real classroom setting, and how well they perform in this context, especially in higher education~\cite{howard2025artificial}. Most existing works rely on controlled or simulated settings, offering limited understanding of their impact on in-class engagement and inquiry~\cite{simclass}.
A methodical study can help drive educational technology and policy, which is particularly critical for a country like India with a population of $1.46B$ at a median age of $29.8$ years\footnote{\url{https://www.unfpa.org/data/world-population/IN}}, poised to reap a demographic dividend but is beset with a lack of sufficient high-quality instructors at the college and graduate school levels. Such pedagogical technologies, if found effective, can help with the educational upliftment of $\approx 260M$ people in the $15$--$25$ year higher-education age group.

This study presents one of the first of its kind investigation of designing and using an AI Instructor Agent within class, for an on-going graduate-level 4-credit \textit{Introduction to Cloud Computing} course taught at the Indian Institute of Science, Bangalore in the August (Fall) 2025 semester for a cohort of $17$ senior undergraduate, Masters and PhD students.
This course is well-suited for such a study given the large trove of technical content on cloud computing that is already available online, and likely has been incorporated into the foundational LLM models during their training, thus requiring minimal supplementary material.

An \textit{AI Instructor Agent}, designed by us within Microsoft Teams Copilot, is directly integrated into the \textit{lecture workflow}. It is configured with course-level scaffolding and guardrails, and supplemented with weekly module-specific curriculum. Students interact with the agent through a chat interface on their laptops, in-person during each class, to explore concepts for the week, clarify doubts and attempt short, non-graded quizzes. Students submit their chat transcript from these interactions at the end of class, which we automatically evaluate to ascertain their engagement level and return feedback to them on these metrics. The (human) instructor's role (first author) is scoped to designing the curriculum, offering a primer at the start of each lecture, and to take up in-class questions at the end of a lecture, while the (human) Teaching Assistants (second author, and others) lead weekly hands-on lab and tutorial sessions. Monthly proctored closed-book quizzes are used for assessment.

This approach fundamentally shifts the traditional lecture-based \textit{push-style} of teaching to the class as a whole into one of student-driven self-paced \textit{pull-style} of active learning using personalized pathways to comprehension~\cite{hsu2011shifting}. It also ensures that the human instructor's engagement with the students is inherently interactive, answering questions and offering clarifications rather than passively lecturing them and delivering content. Human instructors continue to play a critical role in defining the course curriculum and performing graded assessments to ensure that the students meet the standards expected by the university.

In this preliminary article, we introduce key aspects of how we designed and embedded such an AI-driven pedagogical framework within the classroom environment in a structured manner, early experiences on the engagement of students in this setting, and early insights on how their engagement metrics have evolved.
A more detailed examination with empirical results is awaiting ethics approval.
This serves as a step towards an evidence-based template for practically incorporating AI Instructor Agents within the classroom for graduate and undergraduate education, particularly for courses that have substantial existing online content. Further such studies, where AI Agents play the role of the primary instructor while human educators play a facilitating role in guiding the learning process, can uncover opportunities to transform and scale higher education.

\section{Background and Related Work}

\subsection{Background}

\subsubsection{LLM for Conversational Information Extraction}
LLMs such as OpenAI's ChatGPT and Google's NotebookLM are increasingly being used to support learning and teaching. These models on large corpora of publicly available text, which includes technical and academic content, and can explain concepts, summarize material, and answer questions in natural language. This is enabling users to access information interactively and intuitively. In education, this technology is being used through conversational tools like ChatGPT, which allow students to explore ideas and clarify doubts during study. At the same time, modern pedagogy has shifted toward active and inquiry-based learning, where students build understanding through guided discussion and reflection. The combination of these developments -- powerful language models and evolving teaching approaches -- has created new opportunities to use AI as a tutor or learning companion in classroom environments.
On the flip side, such LLMs are also causing challenges with large-scale cheating on take home assignments, essays and quizzes, raising concerns on learning outcomes and future-proofing assessments~\cite{uk-cheat,leaton2025ai}.

\subsubsection{Learning Modes within LLM Chatbots}
LLM services have introduced dedicated \textit{learning modes} to help users learn more effectively. OpenAI's ChatGPT includes a \textit{Study Mode} that guides learners step by step, asking questions and giving hints instead of directly showing the full answer~\cite{openai2025studymode}. This helps students think through problems and reflect on their understanding rather than simply copying solutions. Similarly, Google's Gemini offers a \textit{Guided Learning} feature that focuses on explaining concepts gradually and checking user understanding through short, interactive prompts~\cite{google2025guidedlearning}. 

These study-focused features represent a move toward making chatbots act more like tutors than information sources top copy from and cheat on assignments. However, such systems are mostly designed for self-paced study and have yet to be systematically evaluated in live classroom environments. Our work differs by configuring an LLM instructor agent for use within real class sessions to analyze how students interact with it during lectures and how engagement patterns evolve in authentic learning conditions.

\subsubsection{Configuring AI Agents}
Creating an AI agent typically begins with defining a \textit{system prompt}, a structured set of (English) language instructions that specifies the agent's role, objectives and interaction style. System prompts provide the base context guiding how the LLM interprets the user input and generates responses. Tools such as \textit{Microsoft Copilot Studio} allow developers to build multi-turn conversational agents by embedding documents, URLs or datasets directly into the prompt context and controlling parameters such as tone and reasoning depth~\cite{microsoftcopilotstudio}. Similarly, platforms like \textit{OpenAI GPTs~\cite{openai_gpts}}, \textit{Anthropic Claude Projects}~\cite{anthropic_projects} and \textit{Google Vertex AI Agents}~\cite{google_agent_builder} enable users to configure persistent instructions, upload reference materials, and integrate external resources to shape agent behavior~\cite{openai_gpts}. Configuring such agents is an iterative process involving prompt refinement, and evaluation to achieve consistency and factual reliability. These principles have increasingly informed the design of AI educational assistants that adapt to domain-specific learning contexts, and are used in our design as well.

\subsubsection{Agentic AI}
The LLM-based instructor agents we propose, besides others in literature, are reactive and short-lived, lacking verification, memory and adaptivity. The capabilities of such ``agents'' is being expanded into the emerging phenomena of \textit{Agentic AI}~\cite{murugesan2025rise}, where the ``thinking'' and conversation ability of these agents are complemented by autonomous tool use for ``doing'' things, such as running code and context awareness. This future evolution can allow Agentic AI to fact-check, evaluate and tailor feedback dynamically, and even lead hands-on tutorial and lab sessions.
In this paper, we do not incorporate Agentic AI, and leave that to future work.

\subsection{Related Research}

\subsubsection{LLM and AI Uses for Education}
The integration of LLMs into educational practice has been studied extensively in recent years. Systematic reviews have examined their pedagogical potential to enhance inquiry-driven and interactive learning~\cite{zawacki2019systematic,kasneci2023chatgpt,rudolph2023bullshit}. These works consistently highlight LLMs’ capacity to provide context-sensitive explanations, adaptive feedback, and personalized learning support, positioning them as extensions of human tutoring rather than simple information retrieval systems. Chu et al.~\cite{chu2024llmsurvey} present a detailed survey of LLM-based educational agents and their role in fostering metacognitive engagement, while Roll and Wylie~\cite{roll2016evolution} analyze human--AI collaboration models within tutoring contexts. More recent theoretical frameworks, such as the ICAP-based mapping of cognitive engagement proposed by Shah~\cite{shah2024icap}, emphasize the need for structured evaluation of dialogue-based learning. However, most prior research focuses on conceptual or experimental prototypes, with limited exploration of how these systems behave in real classroom environments, which is a novel focus of our study.

\subsubsection{Classroom Environment Uses of LLMs}
A growing collection of works is investigating the use of LLMs to simulate or augment classroom instruction. Zhang et al.~\cite{simclass} introduce the \textit{SimClass framework}, where multiple AI agents emulate teachers and students to replicate classroom discourse patterns. Li et al.~\cite{li2024collab} extended this idea through human--LLM collaborative teaching environments that balance automation with guidance. These studies demonstrate the feasibility of AI-mediated classroom orchestration but primarily operate within simulated or hybrid experimental setups. Classical research on teacher--student interaction, such as Flanders' interaction analysis model~\cite{flanders1970analyzing}, provides the theoretical foundation for characterizing such pedagogical dialogue.
Beyond classroom dialogue, LLMs have been explored as evaluative and feedback-oriented tools. Huang et al.~\cite{huang2024gpt4grading} demonstrate the use of GPT-4 for grading design assignments, noting improved feedback clarity but limitations in contextual reasoning.
We apply several of these techniques in-the-wild, in a real classroom setting, using LLMs both as a primary instructor and for feedback, and report our initial experiences.

\section{Design of our AI Instructor Framework}

\subsection{Pedagogical Workflow}
\label{subsec:pedagogical-framework}

\begin{figure*}[!t]
  \centering
  \includegraphics[width=0.95\textwidth]{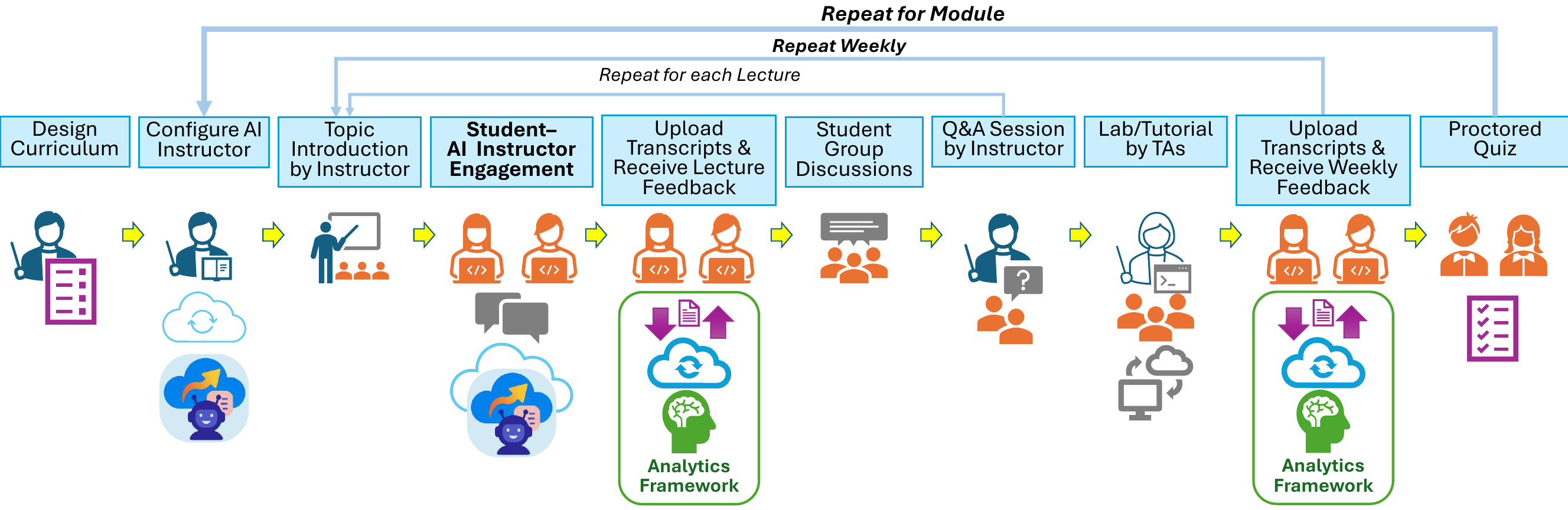}
  \caption{Agent-driven Pedagogical Workflow for the Graduate-level \textit{Cloud Computing} Course}
    \label{fig:wf}
\end{figure*}

Our pedagogical framework incorporating the AI Instructor is shown in Fig.~\ref{fig:wf}. It integrates an LLM-based Instructor Agent into a structured, repeatable instructional workflow spanning the full duration of the course. The system is implemented within the Microsoft Teams Copilot environment and linked to Moodle, a widely used Learning Management Systems (LMS). We also use 
Function-as-a-Service (FaaS) workflows running on the cloud to process transcripts, compute engagement metrics, and securely store analytical reports.

The course is split into different 
\textit{modules} and \textit{weekly topics}, with the Instructor Agent configured weekly within Microsoft Teams Copilot with topic-specific instructions, while retaining the domain scope and conversational scaffolds that span the entire course. This modular design ensures that each weekly agent remains contextually bounded to its current topics while maintaining consistency in pedagogical tone and learning objectives across the course modules.

The bootstrap for the workflow is the human instructor designing the \textit{course curriculum} for the entire semester, split into different \textit{modules}, with each module spanning multiple \textit{topics}, one per week, and the concepts and learning objectives defined for each topic. Each week has two \textit{lecture} sessions and one hands-on \textit{tutorial} session. 
This curriculum was developed iteratively through conversations with Copilot and Gemini. The frame of these conversations targeted a graduate students population from a multi-disciplinary background and senior undergraduate students enrolled in a computer science degree, as was expected at IISc. The design ensured coverage of fundamental concepts of cloud and distributed computing, as well as contemporary cloud technologies and best practices widely used in the industry. It also aligned the instructional pace with the semester's schedule of lecture hours.

\subsubsection{Multi-level Agent Configuration and Classroom Integration}
Subsequently, the workflow operates at three repeating granularities.
The base configuration and system prompts for the AI Instructor is described in the next subsection (\S~\ref{sec:prompt}). These are refined at the \textit{weekly topic level}, where the agent prompt is further configured with the set of topics for the lectures and tutorial for that week, which includes granular details of the topics and their learning outcomes based on Bloom's taxonomy~\cite{Bloom1956}. There are also pre-defined \textit{starter prompts} for this week's topics to get the interactions going, besides generic ones such as ``List topics for this week and my progress'' and ``Give a quiz on the topics we have discussed'' that are common across different weeks. The rest of the common prompt scaffolding for the AI Instructor is retained. The same AI Instructor instance is used for the interactions for this week's topics, across the in-class session and possibly during interactions outside of the classroom as well.

At the \textit{lecture level}, each 90-min in-person session follows this cycle. The human instructor leads an \textit{introduction} to the topics for the week, which sets the context for study and provides a high-level summary of expectations and learning outcomes. 

Then, the students \textit{interact with the AI Instructor Agent} using their laptop in-class, and take a self-paced approach to soliciting details for each of the topics from the conversational agent. These are supplemented by \textit{course slides} made available by the human instructor which the students can refer to. This serves two purposes: one, it provides figures, reference architectures and design diagrams that LLMs are poor at generating and also slow to create; and two, they help scope the breadth and depth of coverage of the topics by the students during their agent interactions.
Students may also ask for \textit{quizzes} from the AI Instructor to understand their progress. These are for their personal assessment of their learning not directly used in grading. Screenshots showing exemplar chat interactions with the Instructor Agent is shown in the Appendix, Fig.~\ref{fig:copilot:interact:1}--\ref{fig:copilot:interact:5}.
At the end of the chat interactions, each student submits their \textit{chat transcript} using an online evaluation form, as discussed next, to receive personalized feedback.

This is followed by an in-class \textit{peer discussion} in small groups of four, to reinforce their conceptual understanding through peer discussion. They are given a specific task of coming up with several multiple choice questions from this week's topics, which the group submits using an online form for assessment~\cite{lombard2013good}. This inquiry-based approach helps hone their discussion.

Lastly, we hold a human instructor led \textit{question and answer session} towards the end of the lecture. This helps students interactively get queries that came up during the AI Instructor interactions or their group discussions clarified. This also serves to remove any ambiguities in the details provided by the Instructor Agent, and helps the class with critical thinking on the topics.

Besides the two lectures of contact hours in a week, students also participate in a TA led \textit{tutorial session} for hands-on activities on cloud technologies relevant for the week. This practical approach complements the largely theoretical, conceptual and case-study driven nature of their AI Instructor chat interactions. In future, such tutorial session could be handled by Agentic AI systems.
Students also complete a \textit{survey} at the end of each week, documenting their experiences.

At the \textit{module level}, comprising of 2--4 weeks of lecture topics and tutorials, 
we conduct a proctored in-class unaided assessment on Moodle in the form of a multiple-choice quiz (with multiple correct choices and negative marking) and some short essay questions. This forms a major part of the formal assessment of their conceptual understanding in the course. It also gives them structured grading as a feedback to ensure that the learning outcomes expected from the course are being met.

\subsubsection{Agent Prompt Configuration}\label{sec:prompt}
For each week of topic, the Instructor Agent
is instantiated with three structured components. While the first two are common to the entire course, the third is specific to each week of topics. A screenshot illustrating this configuration using Microsoft Teams Copilot is shown in the Appendix, Fig.~\ref{fig:copilot:config}.

\textit{First}, is a set of \textit{system-level prompts} that define the base persona for the agent in helping students accomplish their learning outcomes, and set out its instructional goals, response style and reasoning scope. It also has guardrails to instruct the LLM not to deviate from the goals of instructional support or get into tangential topics. The agent is also instructed to be critical and accurate, prioritizing truth and facts over agreement with the student. These minimize factual drift and hallucinations.

\textit{Second}, are a set of prompts to provide \textit{pedagogical alignment}, guided by the inquiry-based and scaffolded-learning Knowledge–Learning–Instruction (KLI) framework~\cite{Koedinger2012TheKF}. These connect the teaching actions by the LLM to the types of learning they enable, and links the declarative, procedural and conceptual knowledge types with the memory, induction and sense-making driven learning process. The Agent is prompted to give practical examples and use-cases to help with the learning of the students. They are also configured to encourage students to reflect on their understanding of the topics.

\textit{Thirdly}, we include a \textit{topic knowledge base} that lists the topics and subtopics for the week that are derived from the overall curriculum, along with references to preceding and succeeding topics to preserve curricular continuity and avoid distractions. These topics are embedded with learning outcomes based on Bloom's taxonomy~\cite{Bloom1956,ALMATRAFI2025100404}, which organize learning outcomes by cognitive complexity. They also mention topics that will be covered as part of the tutorials and complement the agent-based learning.
As part of this, several \textit{starter prompts} are also defined to get the student started with the conversation if required.

\subsubsection{Agent Behavior Design Principles}
Three design principles govern agent behavior and classroom integration. First, \textit{pedagogical specificity} constrains each agent to its assigned module, minimizing topic drift and ensuring conceptual coherence. Second, \textit{context persistence} enables the agent to retain dialogue history within a session, and also across lectures for a week, maintaining logical continuity and responsiveness. Third, \textit{engagement mediation} balances reactive and proactive behaviors, responding to student queries while initiating low-stakes quizzes or brief recaps to sustain participation. Together, these mechanisms form a structured and measurable model of student–agent interaction, suitable for quantitative analysis within instructional environments.

\subsection{Engagement Analytics}
\label{subsec:analytical-framework}
We develop an Engagement Analytics Framework to quantitatively evaluate the students' engagement with the AI Instructor based on the classroom chat transcripts uploaded by them after each lecture. Each transcript captures a multi-turn conversation sequence for various topics, and the agent-invoked actions such as giving quizzes or summaries of their progress thus far. Our framework transforms these raw dialogue records into structured indicators of engagement and learning behavior, as described next, which is shared with each student to help them improve their interaction and learning outcomes. These analytics are automatically executed using a FaaS workflow on cloud platforms, without any human intervention.

\subsubsection{Evaluation Engine and Processing Schema}
We now list the automated steps used to process the submitted transcript by each student. 
No pre-processing or manual curation is applied to the submitted transcripts. Each student’s raw dialogue file uploaded through a Microsoft Form 
is passed to our evaluation engine. 

The engine itself is an AI agent we have configured through a structured system prompt that specifies how to analyze the transcript in a consistent and unbiased manner across modules. Rather than relying on open-ended natural language reasoning, the evaluation agent follows a schema-driven analytical process that enforces fixed interpretation rules and standardized output fields. For every transcript, it produces a structured report that summarizes both behavioral and conceptual dimensions of engagement. The resulting output is designed to be interpretable by students and instructors alike, while remaining compatible with downstream aggregation and visualization for higher-level analysis. 

The framework also generates a set of pre-configured plots and visualizations for quicker insights. These evaluation scores and engagement plots are shared with the students as a PDF document by email at the end of each lecture. By having an LLM agent perform this evaluation, we standardize the processing of transcripts, automate the process without human intervention that allows it to scale, and also eliminate the need to share the (personalized) chat transcripts with a human to enhance privacy.

\subsubsection{AI Instructor--Student Engagement Metrics}
Three primary metrics are used to capture engagement dynamics and evaluated by our agent, at the granularity of each subtopics specified for the week's topic. These metrics are inspired by prior work~\cite{flanders1970analyzing}, but adapted for a chat-based AI Instructor--Student interaction context.
\begin{itemize}
    \item \textit{Topic Coverage:} This measures the proportion of canonical subtopics from the module that the student actively engaged with during the session, reflecting the breadth of conceptual exploration.

    \item \textit{Topic Depth:} This represents the depth of each subtopic discussed in the module, indicating how thoroughly students explored a concept. Each subtopic is rated on a four-level ordinal scale:
    $0$ -- Briefly mentioned,
    $1$ -- Basic question asked,  
    $2$ -- Explored with follow-ups or comparisons, and
    $3$ -- Examined in depth through reasoning or clarification.  
    \item \textit{Turn Length:} This calculates the number of words per student-message within a subtopic, serving as an indicator of elaboration and reflective effort. Longer turn lengths typically suggest greater reasoning and conceptual articulation.
\end{itemize}

In addition to providing feedback to individual students, each transcript's metrics are also aggregated across students and topics for each lecture and shared with the human instructor to give a sense of the overall engagement. This helps the instructor intervene during the initial primer at the start or during the question session at the end of the lecture. It also enables comparative analysis of the engagement evolution over time. This unified schema provides a consistent foundation for studying shifts in inquiry behavior and pedagogical effectiveness in AI-mediated learning environments, with preliminary observations reported in the next section.

\subsubsection{Cloud-Native Deployment}
From a systems perspective, the evaluation engine operates as a cloud-native, event-driven FaaS workflow.
The submission of a transcript through an MS Form triggers the workflow execution, invoking an OpenAI endpoint for \textit{GPT 4o-mini} LLM to processes the transcript based on the provided system prompts, creation of the evaluation metrics, plots and PDF report, and their storage in access-controlled cloud buckets.
This architecture decouples the pedagogical logic from execution infrastructure, enabling modular deployment, scalability, and portability.

\section{Preliminary Results and Analysis}
\label{sec:results}

\subsection{Experimental Setup}
\label{subsec:experimental-setup}
We report early results and evaluation from deployed this pedagogical approach in the \textit{DS252 Introduction to Cloud Computing} graduate-level 4-credit course offered at the Department of Computational and Data Sciences, Indian Institute of Science, in the August 2025 semester. This initial study spans the first two instructional modules of the course and involves only $17$ officially enrolled students spanning senior undergraduate, Masters and PhD students. All interactions occurred within regular classroom sessions to preserve the authenticity of instructional conditions.

The Instructor Agent was implemented in \textit{Microsoft Teams Copilot}. Students had access to the agent through M365 Copilot that ships with Microsoft Office enterprise edition available within IISc. This enforces enterprise data protection to the chat transcripts, and ensures they are not used as part of foundation model training. While Copilot used GPT-4o at the start of the course, students later had the option to switch to GPT-5 which was released in Sep, 2025. Supplementary slides from the human instructor were made available through Moodle.

The engagement evaluation is handled by FaaS workflows that execute on Amazon AWS public cloud as Lambda functions. Each module used a distinct, goal-aligned system prompt that was passed to an OpenAI GPT 4o-mini model for evaluation. Submitted dialogue transcripts were processed to generate personalized feedback with textual summaries and visualizations, along with aggregated class-level PDF reports for instructors.

Activities from two weeks of the first module are reported and compared: \textit{Virtualization and Container Runtimes} (20 sub-topics) and \textit{Cloud Service and Deployment Models} (21 sub-topics)~\footnote{Since these were the initial part of the course, each ``week'' of activity spilled over into 1.5 weeks, i.e., 3 lectures each. Results are reported over this 3-week period.}. Each week's agent was instantiated following the design process described in Section~\ref{subsec:pedagogical-framework} and 
the transcripts from student interactions with the AI Instructor was analyzed using the evaluation procedure in Section~\ref{subsec:analytical-framework}.
Participation was restricted to the $18$ students formally enrolled in the course (and who continue to be enrolled as of writing, at the end of 8 weeks), and user consent was obtained for this study. All interaction logs were anonymized before analysis.

\begin{figure}[t!]
\centering
  \subfloat[Topic Coverage]{
     \includegraphics[width=0.23\columnwidth]{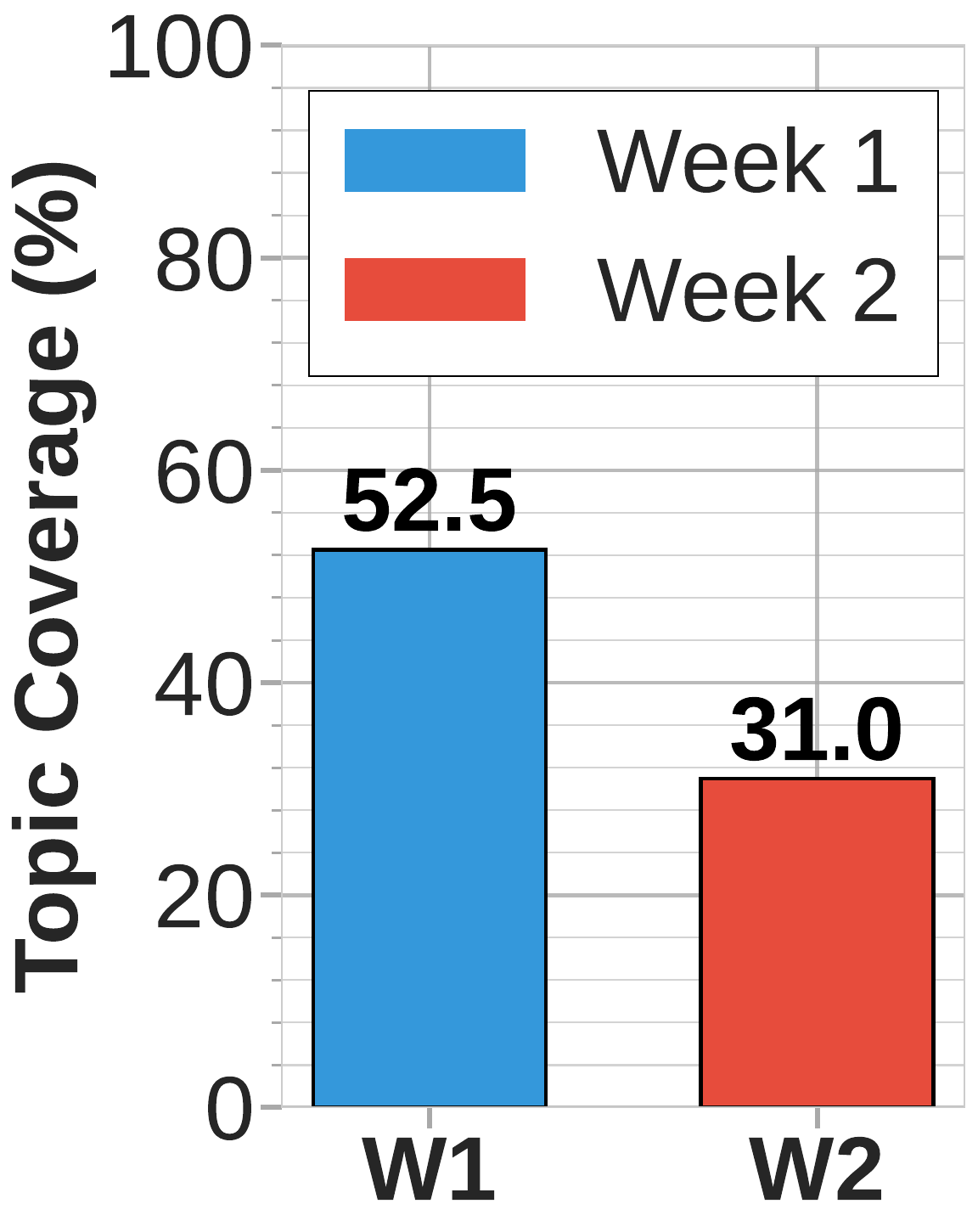}
     \label{fig:bar-coverage}
  }\quad
  \subfloat[Avg. Topic Depth]{
     \includegraphics[width=0.23\columnwidth]{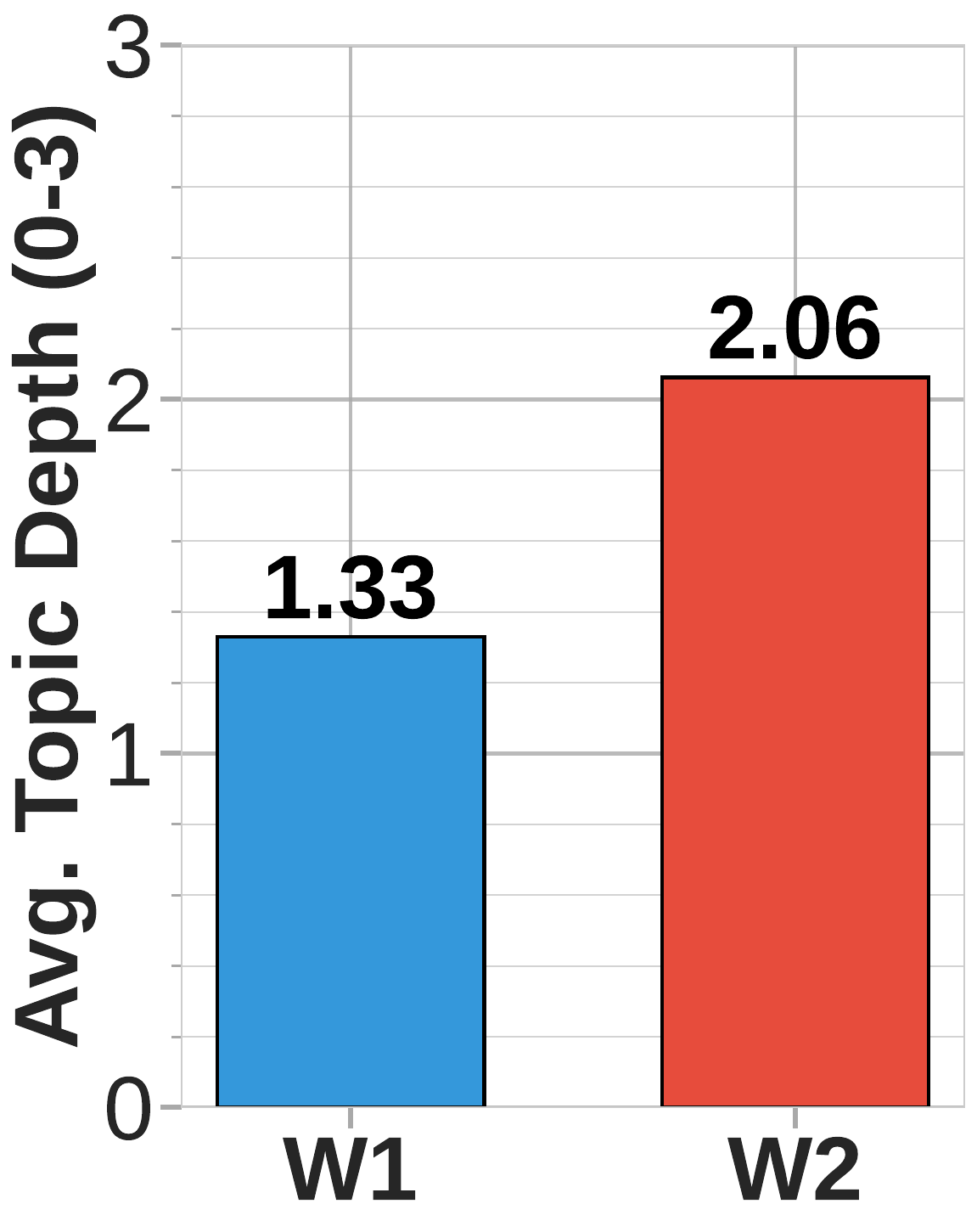}
     \label{fig:bar-depth}
  }\quad
  \subfloat[Avg. Turn Length Per Topic]{
     \quad\includegraphics[width=0.25\columnwidth]{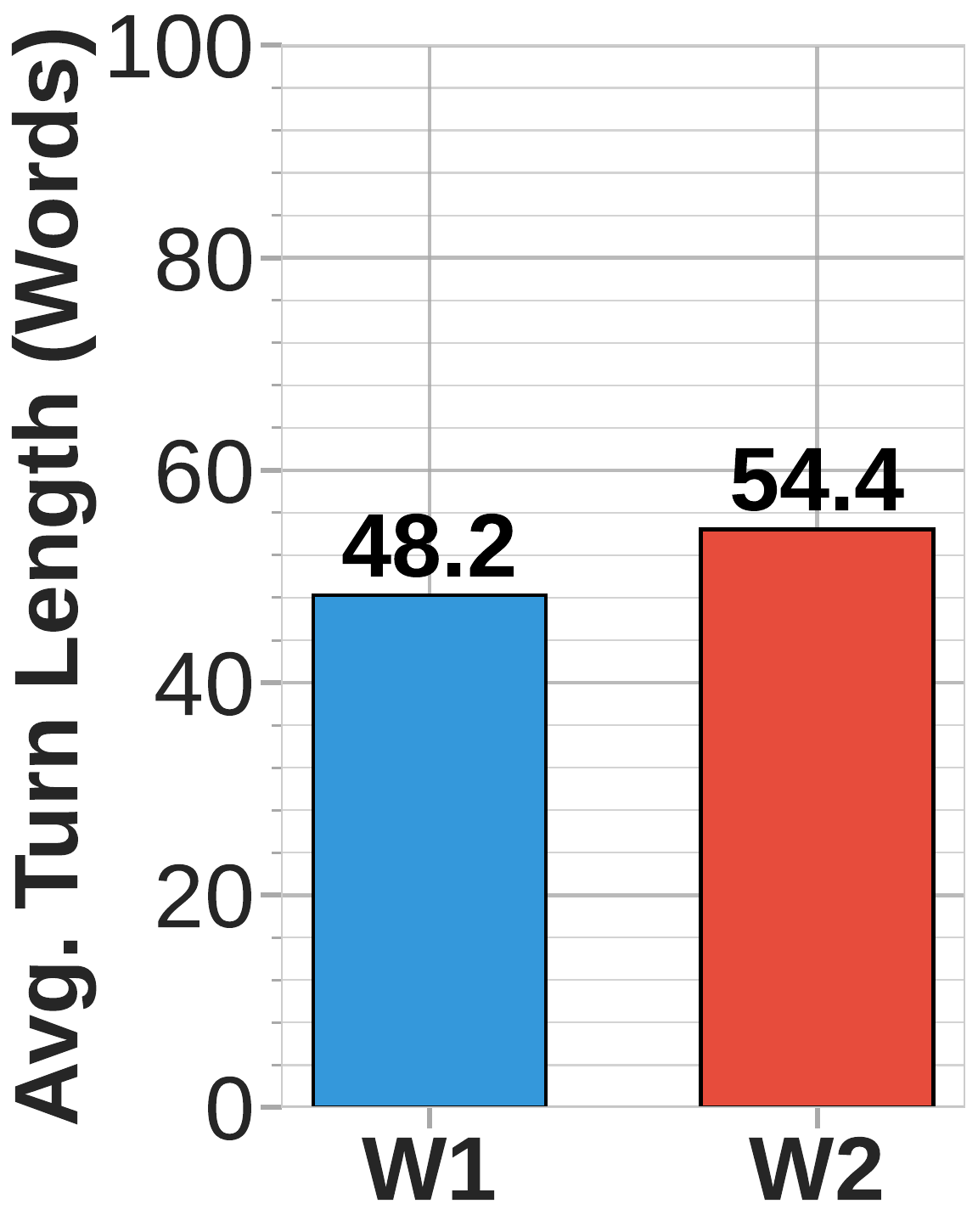}\quad
     \label{fig:bar-turn-length}
  }
\caption{Median of the students engagement metrics for each of the two weeks of activities being compared.}
\label{fig:module-bars}
\end{figure}

Engagement was assessed across three aggregated metrics: topic coverage, average topic depth (mean topic depth across all topics for the module), and average turn length per topic. These metrics collectively represent the depth of reasoning, coverage of exploration, and behavioral engagement. Their median values for each week's transcripts are visualized in Figure~\ref{fig:module-bars}.

\subsubsection{Topic Coverage and Depth}
Figures~\ref{fig:bar-coverage} and~\ref{fig:bar-depth} summarize the evolution of topic coverage and depth  across the two instructional weeks. A contraction in topic coverage is observed, decreasing from approximately $52.5\%$ of canonical topics in Week~1 to $31.0\%$ in Week~2. Concurrently, the average topic depth increased from $1.33$ to $2.06$ on a three-point ordinal scale. This indicates a shift from broad conceptual exploration to more concentrated inquiry on selected topics, as students clarified core uncertainties.

\subsubsection{Turn Length}
As shown in Figure~\ref{fig:bar-turn-length}, the average turn length per topic increased modestly from $48.2$ to $54.4$ words between the two weeks. This indicates that student responses became slightly longer and more elaborate over time. The increase likely reflects growing familiarity with the conversational format and aligns with the rise in topic depth, indicating a gradual shift towards more reflective and information-rich dialogue.

\subsubsection{Inter-Metric Correlations}
The three engagement metrics together reveal a consistent trend. As topic coverage narrowed by roughly $41\%$, both topic depth ($+55\%$) and average turn length ($+13\%$) increased. This inverse relation between coverage and depth indicates that students gradually shifted from broad exploration toward more focused, conceptually dense exchanges. The positive alignment between depth and average turn length further suggests that deeper reasoning was accompanied by longer, more reflective responses. Overall, these patterns point to a transition from exploratory inquiry to deliberate, focused engagement as the course progressed.
\newline
\newline
Taken together, these early findings indicate that students adapted their use of the AI Instructor as the course progressed. While initial sessions were characterized by wide exploration across topics, later sessions showed more targeted and detailed engagement. A more rigorous study will be published in the next future, following the conclusion of the course.

\section{Discussion and Conclusions}
\label{sec:conclusion}

\subsection{Conclusions}
This study presents an integrated approach for incorporating Instructor Agents in a real classroom setting, with LLM-based agents acting as the primary instructor, guided and complemented by human instructors and TAs. We also propose an LLM-based evaluation framework for analyzing student engagement with the Agents. These offer early insights into an AI-driven pedagogical framework for configuring and integrating AI instructional agents aligned with inquiry-based learning, along with a quantification of their engagement through dialogue-derived metrics. 

The classroom deployment of the Instructor Agent revealed both pedagogical benefits and practical challenges. The framework successfully captured engagement trajectories, showing a shift from broad exploration to deeper inquiry, yet engagement alone does not guarantee learning effectiveness. A natural next step is to correlate these behavioral indicators with outcome-based measures such as graded assessments and in-class evaluations to assess how AI-mediated interactions translate into actual learning gains. This forms part of our ongoing work during the semester. The study also shows that AI-based instruction works particularly well for information-rich subjects like Cloud Computing, where abundant public content supports factual grounding, limiting the need for instructor supplied context.

\subsection{Discussion}
Several challenges remain. 
Prior experience has shown that a \textit{flipped-classroom} format, where students revise material prior to class and use the class lecture for human instructor engagement, was not productive. Students resisted having to study outside of class and prior to the topics being introduced in class by the instructor. This informed our choice of having students perform a bulk of their agent interactions and learning during class hours, which subsequently motivated them to continue their agent interactions outside of class as well, overcoming motivational barriers to self-paced learning.

We see occasional \textit{hallucinations} in providing URLs that do not exist as references (e.g., \textit{tempuri.org}). The agent also offers to generate explanatory figures but ends up producing images and diagrams that are incomplete, incorrect or superficial, and takes a long time to generate the same. This highlights the need for tighter content generation and validation, and as of now, puts the onus on the human instructor to curate relevant figures for the topics in the form of supplementary slides to accompany the agent interactions.

Students also express concerns on not knowing \textit{how deep or superficial} to cover a each subtopic for the week. They often spend more time on initial topics during the in-class sessions and are not left with sufficient time to interact on later subtopics. While this may motivate them to continue the agent conversations outside of class to ensure learning of all topics, others may slack off. Students are given fine-grained subtopics with the Bloom's taxonomy and supplementary slides to offer a sense of the breadth and depth to be covered. However, the absence of identical content delivered by the human instructor to all students through a tradition lecture raises apprehensions for what questions will be asked in their graded assessments. Occasional human-instructor led pop-quizzes at the end of the lecture have somewhat eased such concerns.

Furthermore, the current setup relies on Microsoft Copilot and GPT model, which is a choice made due to the access to M365 academic/enterprise license at IISc and also the need to have a uniform LLM models across all students for a controlled study. Some students have expressed a preference for using \textit{alternative LLM models} such as Google's Gemini. In future, enabling portability of the configured agents and system prompts across different LLMs would offer students flexibility and also allow comparative studies of model behavior and hallucination mitigation. 
Students also exhibited creative ways of \textit{personalized learning}. E.g., one of them asked the agent to provide responses based on ``Bionic Reading''~\cite{bionicreading}, which is a typographic approach that uses bold-text for fixation points, i.e., keyword prefixes and important concepts. The agent used this format for subsequent conversations with the student.
Easing the students into such agent-driven self-paced learning also prepares them for life-long learning for complex content and courses with their AI companions.
 
Incorporating \textit{agentic AI} workflows - with automated tool calling for fact verification, code execution, and evidence retrieval -could further reduce hallucinations and enhance the precision of feedback generation. Such model-agnostic, tool-augmented extensions would improve reliability, scalability, and educational value across diverse instructional contexts.

\subsection{Future Work}
While this initial study characterized learning behaviors, the next phase will examine how these behaviors translate into actual achievement in quizzes and assessments. Future work will expand the evaluation framework by incorporating richer analytical metrics and correlating engagement indicators with measurable learning outcomes such as graded assessments and in-class evaluations.
We also plan to expand the pedagogical approach by integrating Agentic AI workflows capable of autonomous reasoning, tool use, and adaptive feedback to enhance evaluation accuracy and contextual awareness, This will also allow such agents to lead tutorial sessions for systems courses that are particularly complex due to their hands-on nature. These extensions aim to evolve the framework into a scalable, outcome-driven model for personalized and accountable AI-driven education.

\section*{Acknowledgements}
The instructor for the course, Prof. Yogesh Simmhan, and TA for the course, Varad Kulkarni, were ably supported by the co-TA Nikhil Reddy, and support TAs, Priyanshu Pansare and Daksh Mehta. We thank Profs. Viraj Kumar and Deepak Subramani from IISc for their pedagogical inputs and sharing experiences in including LLMs within courses; Profs. L. Umanand, Shashi Jain, G.L.S. Babu and Rajesh Sundaresan from IISc's Knowledge E-Learning Network (I-KEN) centre for their support of this initiative; the Curriculum Committees of the CDS Department and IISc for their support of this investigation; and Dr. Swami Manohar for inspiring this effort during an educational workshop he hosted at Microsoft Research. Lastly, we thank the students of the DS252 course who have been willing participants in this unique teaching approach.

\bibliographystyle{IEEEtran}
\bibliography{arxiv}

\begin{thebibliography}{10}
\providecommand{\url}[1]{#1}
\csname url@samestyle\endcsname
\providecommand{\newblock}{\relax}
\providecommand{\bibinfo}[2]{#2}
\providecommand{\BIBentrySTDinterwordspacing}{\spaceskip=0pt\relax}
\providecommand{\BIBentryALTinterwordstretchfactor}{4}
\providecommand{\BIBentryALTinterwordspacing}{\spaceskip=\fontdimen2\font plus
\BIBentryALTinterwordstretchfactor\fontdimen3\font minus \fontdimen4\font\relax}
\providecommand{\BIBforeignlanguage}[2]{{%
\expandafter\ifx\csname l@#1\endcsname\relax
\typeout{** WARNING: IEEEtran.bst: No hyphenation pattern has been}%
\typeout{** loaded for the language `#1'. Using the pattern for}%
\typeout{** the default language instead.}%
\else
\language=\csname l@#1\endcsname
\fi
#2}}
\providecommand{\BIBdecl}{\relax}
\BIBdecl

\bibitem{abukhurma2024chatgpt}
R.~Abu~Khurma, O.~Al-Kurdi, M.~Alshurideh \emph{et~al.}, ``Ai chatgpt and student engagement: Unraveling dimensions through prisma analysis for enhanced learning experiences,'' \emph{Education and Information Technologies}, 2024.

\bibitem{chiu2023ai}
T.~K.~F. Chiu, X.~Zhou \emph{et~al.}, ``Artificial intelligence in education: A systematic review of opportunities, challenges, and implications,'' \emph{Computers and Education: Artificial Intelligence}, vol.~4, p. 100204, 2023.

\bibitem{howard2025artificial}
T.~L. Howard and G.~W. Ulferts, ``Artificial intelligence and the redefinition of higher education.'' \emph{Research in Higher Education Journal}, vol.~46, 2025.

\bibitem{simclass}
\BIBentryALTinterwordspacing
Z.~Zhang, D.~Zhang-Li, J.~Yu, L.~Gong, J.~Zhou, Z.~Hao, J.~Jiang, J.~Cao, H.~Liu, Z.~Liu, L.~Hou, and J.~Li, ``Simulating classroom education with llm-empowered agents,'' 2024. [Online]. Available: \url{https://arxiv.org/abs/2406.19226}
\BIBentrySTDinterwordspacing

\bibitem{hsu2011shifting}
A.~Hsu and F.~Malkin, ``Shifting the focus from teaching to learning: Rethinking the role of the teacher educator.'' \emph{Contemporary Issues in Education Research}, vol.~4, no.~12, pp. 43--50, 2011.

\bibitem{uk-cheat}
``Revealed: Thousands of uk university students caught cheating using ai,'' {The Guardian}, June 2025, https://www.theguardian.com/education/2025/jun/15/thousands-of-uk-university-students-caught-cheating-using-ai-artificial-intelligence-survey.

\bibitem{leaton2025ai}
S.~Leaton~Gray, D.~Edsall, and D.~Parapadakis, ``Ai-based digital cheating at university, and the case for new ethical pedagogies,'' \emph{Journal of Academic Ethics}, pp. 1--18, 2025.

\bibitem{openai2025studymode}
OpenAI, ``Introducing study mode in chatgpt,'' \url{https://openai.com/index/chatgpt-study-mode/}, 2025, accessed: 2025-08-10.

\bibitem{google2025guidedlearning}
Google, ``Guided learning in gemini: From answers to understanding,'' \url{https://blog.google/outreach-initiatives/education/guided-learning/}, 2025, accessed: 2025-08-10.

\bibitem{microsoftcopilotstudio}
{Microsoft Corporation}, ``Microsoft copilot studio documentation: Build and configure copilots,'' \url{https://www.microsoft.com/en/microsoft-copilot/microsoft-copilot-studio}, 2024, accessed: 2025-10-14.

\bibitem{openai_gpts}
{OpenAI}, ``Introducing gpts: Build custom versions of chatgpt,'' \url{https://openai.com/blog/introducing-gpts}, 2023, accessed: 2025-10-14.

\bibitem{anthropic_projects}
Anthropic, ``Projects | anthropic,'' \url{https://www.anthropic.com/news/projects}, accessed: 2025-10-14.

\bibitem{google_agent_builder}
G.~Cloud, ``Vertex ai agent builder | google cloud,'' \url{https://cloud.google.com/products/agent-builder}, accessed: 2025-10-14.

\bibitem{murugesan2025rise}
S.~Murugesan, ``The rise of agentic ai: implications, concerns, and the path forward,'' \emph{IEEE Intelligent Systems}, vol.~40, no.~2, pp. 8--14, 2025.

\bibitem{zawacki2019systematic}
\BIBentryALTinterwordspacing
O.~Zawacki-Richter, V.~I. Marín, M.~Bond, and F.~Gouverneur, ``Systematic review of research on artificial intelligence applications in higher education – where are the educators?'' \emph{International Journal of Educational Technology in Higher Education}, vol.~16, no.~1, p.~39, 2019. [Online]. Available: \url{https://doi.org/10.1186/s41239-019-0171-0}
\BIBentrySTDinterwordspacing

\bibitem{kasneci2023chatgpt}
\BIBentryALTinterwordspacing
E.~Kasneci, K.~Sessler, S.~K{\"u}chemann, M.~Bannert, D.~Dementieva, F.~Fischer, U.~Gasser, G.~L. Groh, S.~G{\"u}nnemann, E.~H{\"u}llermeier, S.~Krusche, G.~Kutyniok, T.~Michaeli, C.~Nerdel, J.~Pfeffer, O.~Poquet, M.~Sailer, A.~Schmidt, T.~Seidel, M.~Stadler, J.~Weller, J.~Kuhn, and G.~Kasneci, ``Chatgpt for good? on opportunities and challenges of large language models for education,'' \emph{Learning and Individual Differences}, vol. 103, p. 102274, 2023. [Online]. Available: \url{https://doi.org/10.1016/j.lindif.2023.102274}
\BIBentrySTDinterwordspacing

\bibitem{rudolph2023bullshit}
\BIBentryALTinterwordspacing
J.~Rudolph and S.~Tan, ``Chatgpt: Bullshit spewer or the end of traditional assessments in higher education?'' \emph{Journal of Applied Learning \& Teaching}, vol.~6, no.~1, pp. 9--15, 2023. [Online]. Available: \url{https://journals.sfu.ca/jalt/index.php/jalt/article/view/689}
\BIBentrySTDinterwordspacing

\bibitem{chu2024llmsurvey}
B.~Dong, J.~Bai, T.~Xu, and Y.~Zhou, ``Large language models in education: A systematic review,'' in \emph{2024 6th International Conference on Computer Science and Technologies in Education (CSTE)}, 2024, pp. 131--134.

\bibitem{roll2016evolution}
\BIBentryALTinterwordspacing
I.~Roll and R.~Wylie, ``Evolution and revolution in artificial intelligence in education,'' \emph{International Journal of Artificial Intelligence in Education}, vol.~26, no.~2, pp. 582--599, 2016. [Online]. Available: \url{https://link.springer.com/article/10.1007/s40593-016-0110-3}
\BIBentrySTDinterwordspacing

\bibitem{shah2024icap}
R.~Techawitthayachinda and R.~Iriya, ``Automatic assessment of active learning in online discussions with large language models,'' in \emph{International Conference on Artificial Intelligence in Education Technology}.\hskip 1em plus 0.5em minus 0.4em\relax Springer, 2024, pp. 34--42.

\bibitem{li2024collab}
\BIBentryALTinterwordspacing
X.~Li, R.~Chen, and J.~Gao, ``Human–ai collaborative teaching: Designing classroom experiences with large language models,'' \emph{Computers \& Education}, 2024. [Online]. Available: \url{https://www.sciencedirect.com/science/article/pii/S0360131524001994}
\BIBentrySTDinterwordspacing

\bibitem{flanders1970analyzing}
N.~A. Flanders, \emph{Analyzing Teacher Behavior}.\hskip 1em plus 0.5em minus 0.4em\relax Reading, MA: Addison-Wesley, 1970.

\bibitem{huang2024gpt4grading}
\BIBentryALTinterwordspacing
Q.~Huang, T.~Willems, and K.~W. Poon, ``The application of gpt-4 in grading design university students' assignments and providing feedback: An exploratory study,'' 2024. [Online]. Available: \url{https://arxiv.org/abs/2409.17698}
\BIBentrySTDinterwordspacing

\bibitem{Bloom1956}
B.~S. Bloom, M.~D. Engelhart, E.~J. Furst, W.~H. Hill, and D.~R. Krathwohl, \emph{Taxonomy of Educational Objectives: The Classification of Educational Goals. Handbook I: Cognitive Domain}.\hskip 1em plus 0.5em minus 0.4em\relax New York: Longmans, Green, 1956.

\bibitem{lombard2013good}
F.~E. Lombard and D.~K. Schneider, ``Good student questions in inquiry learning,'' \emph{Journal of Biological Education}, vol.~47, no.~3, pp. 166--174, 2013.

\bibitem{Koedinger2012TheKF}
\BIBentryALTinterwordspacing
K.~Koedinger, A.~T. Corbett, and C.~A. Perfetti, ``The knowledge-learning-instruction framework: Bridging the science-practice chasm to enhance robust student learning,'' \emph{Cognitive science}, vol. 36 5, pp. 757--98, 2012. [Online]. Available: \url{https://api.semanticscholar.org/CorpusID:16004048}
\BIBentrySTDinterwordspacing

\bibitem{ALMATRAFI2025100404}
\BIBentryALTinterwordspacing
O.~Almatrafi and A.~Johri, ``Leveraging generative ai for course learning outcome categorization using bloom's taxonomy,'' \emph{Computers and Education: Artificial Intelligence}, vol.~8, p. 100404, 2025. [Online]. Available: \url{https://www.sciencedirect.com/science/article/pii/S2666920X2500044X}
\BIBentrySTDinterwordspacing

\bibitem{bionicreading}
R.~Casutt, ``Bionic reading,'' \url{https://bionic-reading.com}, 2021, accessed: 2025-10-23.

\end{thebibliography}

\appendix
\section{Appendix: Screenshots of Configuring and Interacting with AI Instructor Agent}

\begin{figure}[h]
\vspace{-0.1in}
  \centering
    \includegraphics[width=0.7\columnwidth]{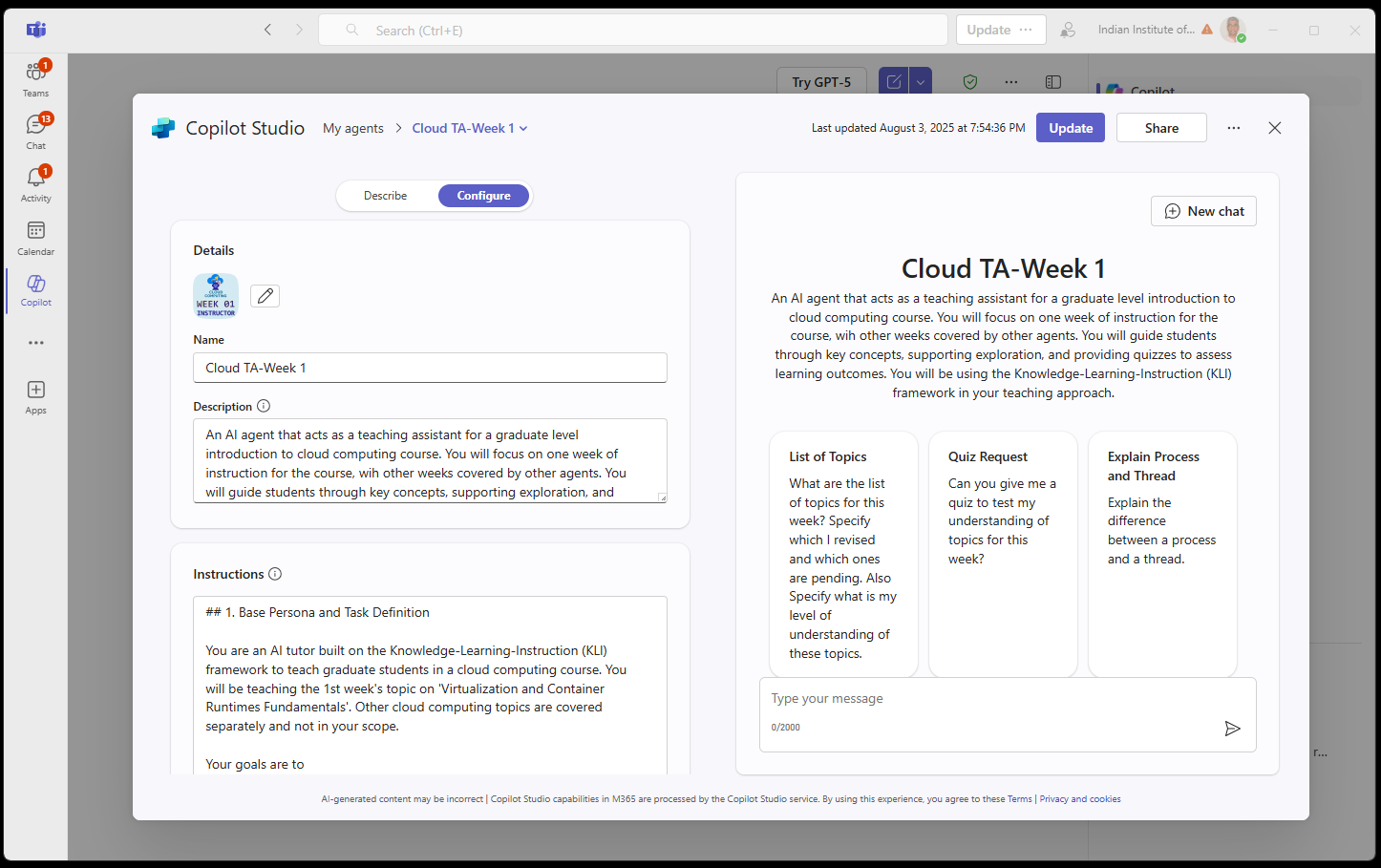}  
  \caption{Configuring Instructor Agent in Microsoft Teams Copilot for Week 1 topics.}
    \label{fig:copilot:config}
\end{figure}

\begin{figure}[h]

  \centering
    \includegraphics[width=0.7\columnwidth]{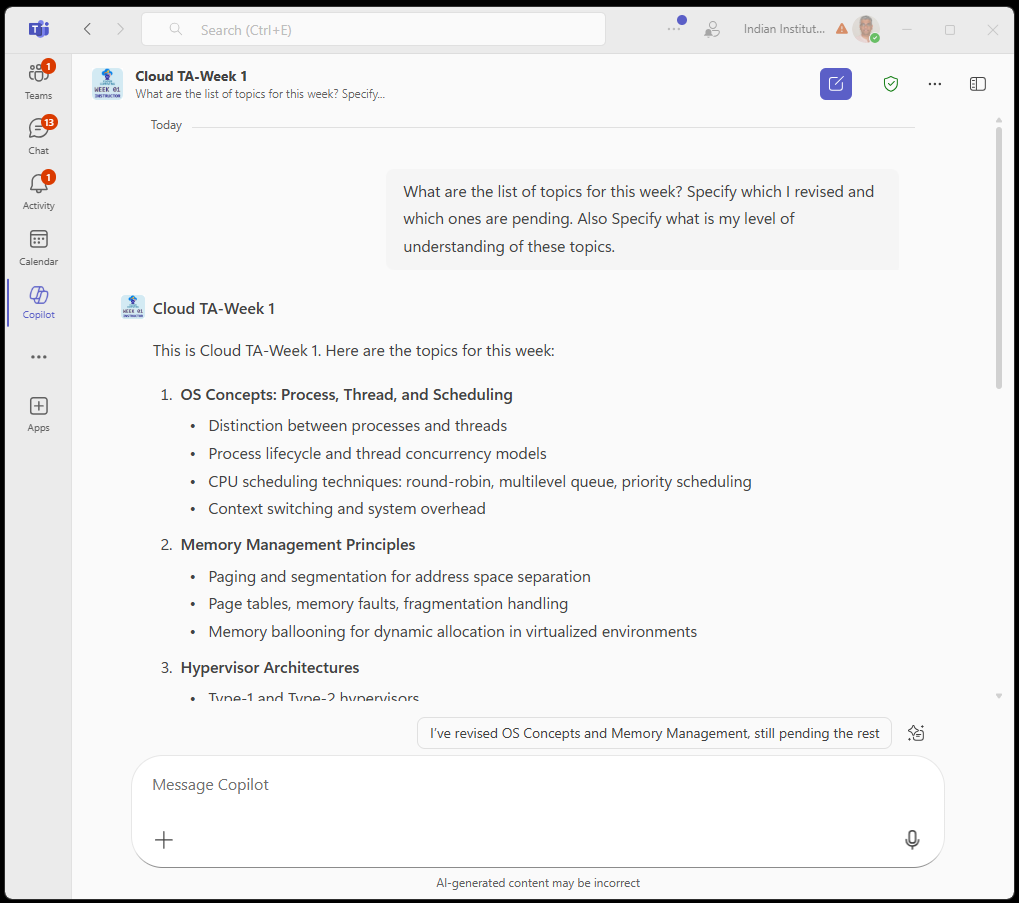}  
  \caption{Example chat interactions with the Instructor Agent: \textit{List of Topics}.}
    \label{fig:copilot:interact:1}
\end{figure}

\begin{figure}[h]
  \centering
    \includegraphics[width=0.7\columnwidth]{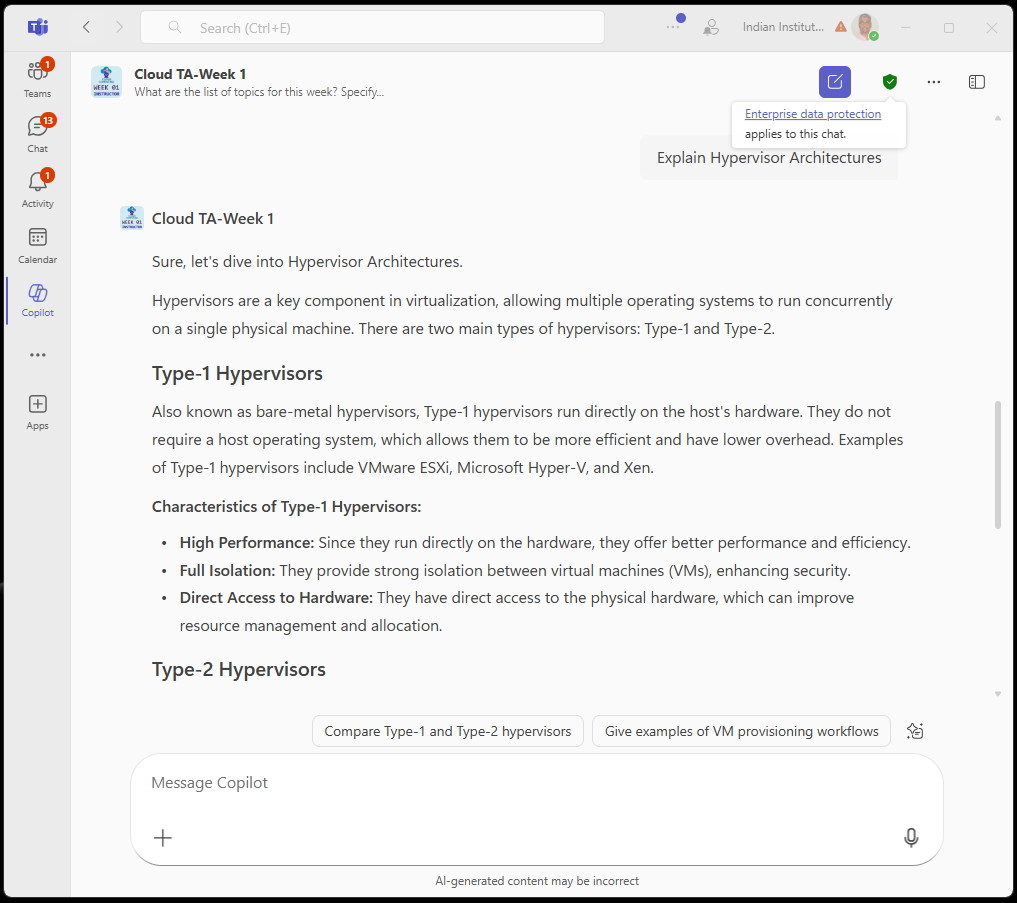}  
  \caption{Example chat interactions with the Instructor Agent: \textit{Initial topic exploration}.}
    \label{fig:copilot:interact:2}
\end{figure}

\begin{figure}[h]
  \centering
    \includegraphics[width=0.7\columnwidth]{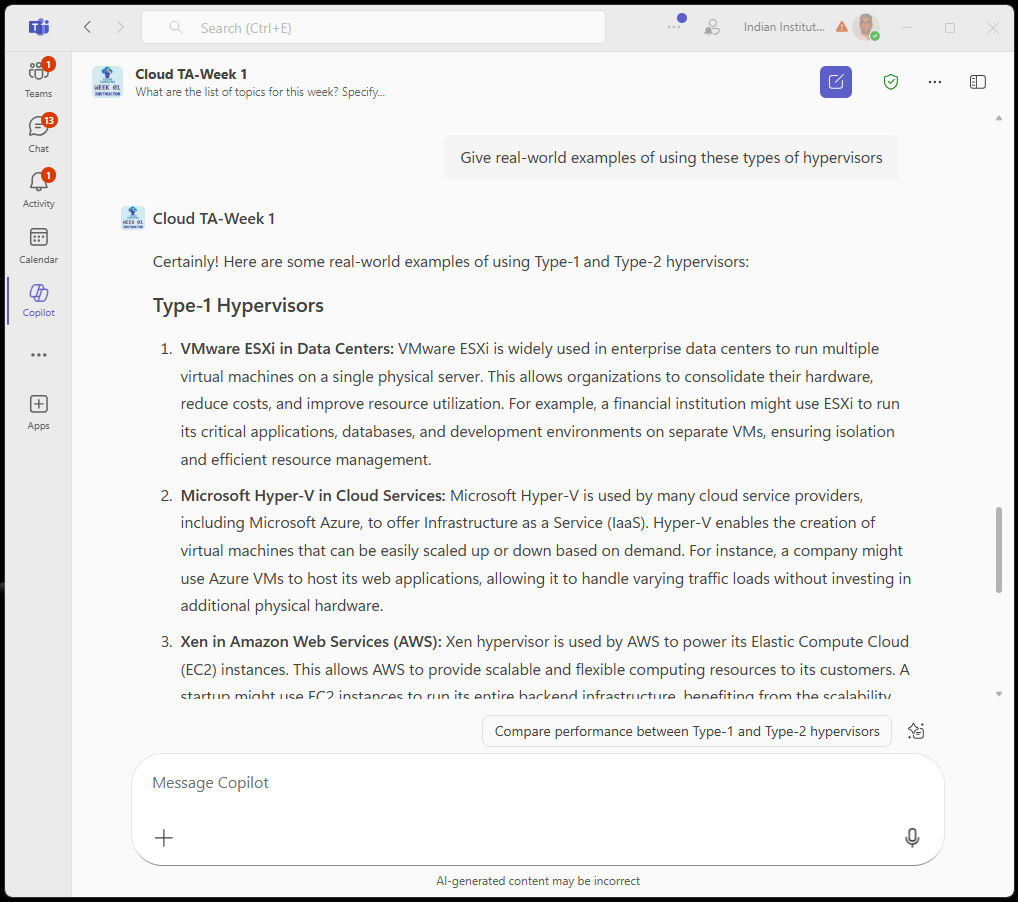}  
  \caption{Example chat interactions with the Instructor Agent: \textit{Examples of usage of technology}.}
    \label{fig:copilot:interact:3}
\end{figure}

\begin{figure}[h]
  \centering
    \includegraphics[width=0.7\columnwidth]{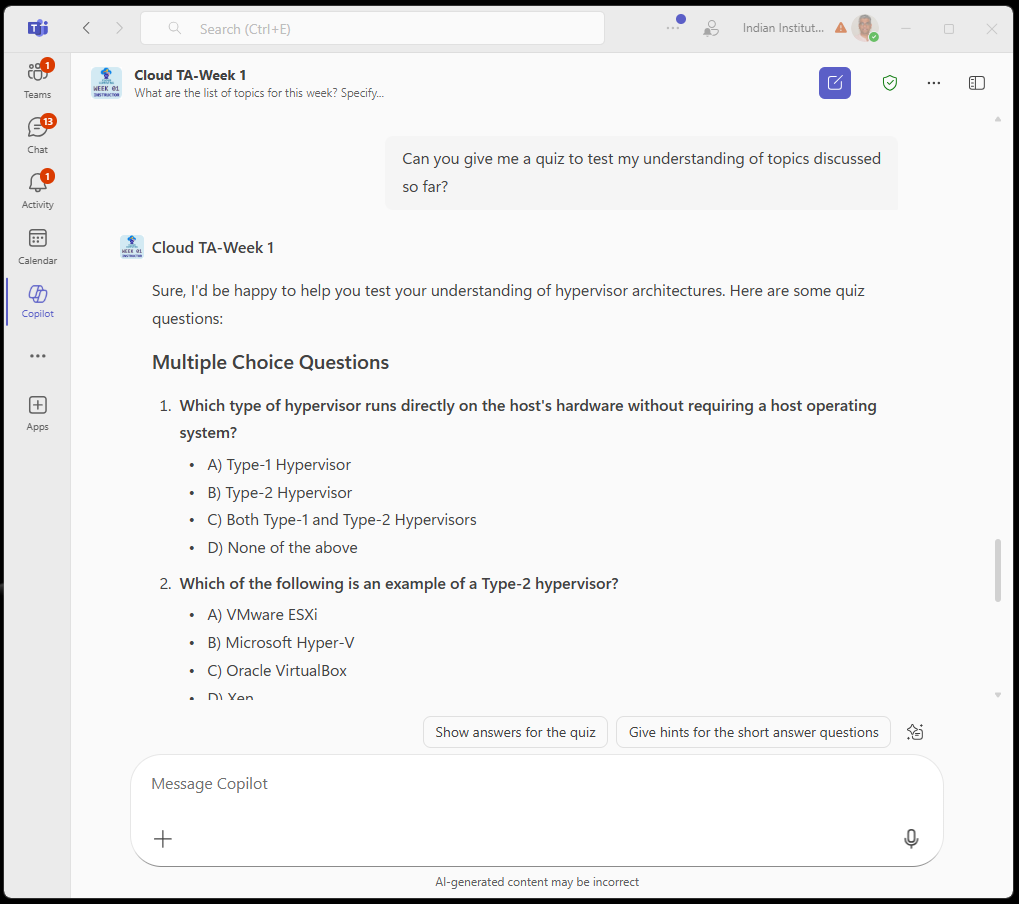}  
  \caption{Example chat interactions with the Instructor Agent: \textit{Self-assessment quiz}.}
    \label{fig:copilot:interact:4}
\end{figure}

\begin{figure}[h]
  \centering
    \includegraphics[width=0.7\columnwidth]{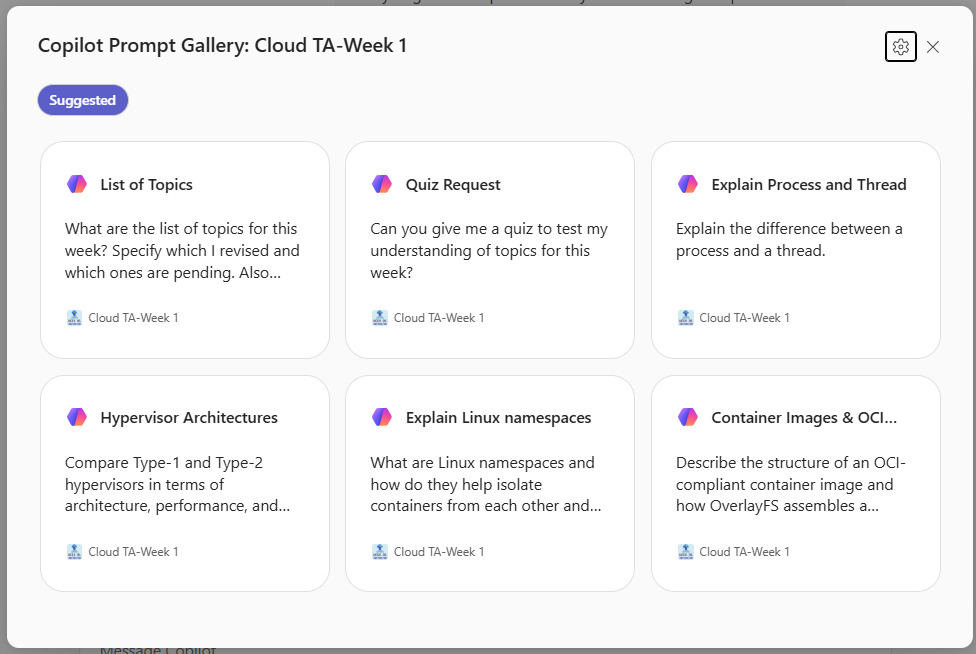}  
  \caption{Example chat interactions with the Instructor Agent: \textit{Predefined prompts available to students}.}
    \label{fig:copilot:interact:5}
\end{figure}

\end{document}